\documentclass[pdflatex,sn-mathphys-num]{sn-jnl}

\usepackage{graphicx}%
\usepackage{multirow}%
\usepackage{amsmath,amssymb,amsfonts}%
\usepackage{amsthm}%
\usepackage{mathrsfs}%
\usepackage[title]{appendix}%
\usepackage{xcolor}%
\usepackage{textcomp}%
\usepackage{manyfoot}%
\usepackage{booktabs}%
\usepackage{algorithm}%
\usepackage{algorithmicx}%
\usepackage{algpseudocode}%
\usepackage{listings}%
\usepackage{tabularx,ragged2e}
\usepackage{booktabs}  
\usepackage{graphicx}
\usepackage{subfig}
\usepackage{array}
\usepackage{color}
\usepackage[figuresright]{rotating}
\newcommand{\bm}[1]{\mbox{\boldmath{$#1$}}}

\begin{document}

\title[Article Title]{How to Find Opinion Leader on the Online Social Network?}


\author[1]{\fnm{Bailu} \sur{Jin}}\email{bailu.jin@cranfield.ac.uk}

\author[1]{\fnm{Mengbang} \sur{Zou}}\email{m.zou@cranfield.ac.uk}

\author[1]{\fnm{Zhuangkun} \sur{Wei}}\email{zhuangkun.wei@cranfield.ac.uk}

\author*[1]{\fnm{Weisi} \sur{Guo}}\email{weisi.guo@cranfield.ac.uk}

\affil[1]{\orgname{Cranfield University}, \orgaddress{\street{College Rd, Cranfield, Wharley End}, \city{Bedford}, \postcode{MK43 0AL}, \country{UK}}}


\abstract{Online social networks (OSNs) provide a platform for individuals to share information, exchange ideas, and build social connections beyond in-person interactions. For a specific topic or community, opinion leaders are individuals who have a significant influence on others' opinions. Detecting opinion leaders and modeling influence dynamics is crucial as they play a vital role in shaping public opinion and driving conversations. Existing research have extensively explored various graph-based and psychology-based methods for detecting opinion leaders, but there is a lack of cross-disciplinary consensus between definitions and methods. For example, node centrality in graph theory does not necessarily align with the opinion leader concepts in social psychology. This review paper aims to address this multi-disciplinary research area by introducing and connecting the diverse methodologies for identifying influential nodes. The key novelty is to review connections and cross-compare different multi-disciplinary approaches that have origins in: social theory, graph theory, compressed sensing theory, and control theory. Our first contribution is to develop cross-disciplinary discussion on how they tell a different tale of networked influence. Our second contribution is to propose trans-disciplinary research method on embedding socio-physical influence models into graph signal analysis. We showcase inter- and trans-disciplinary methods through a Twitter case study to compare their performance and elucidate the research progression with relation to psychology theory. We hope the comparative analysis can inspire further research in this cross-disciplinary area.}


\maketitle



\section{Introduction}

As far back as the 1940s, Paul F. Lazarsfeld, Bernard Berelson, and Hazel Gaudet conducted social influence experiments to understand the social network opinion dynamics towards a topic \cite{lazarsfeld1968people}. As a part of their research, \textit{opinion leaders} were defined as individuals with a significant impact on the opinions, attitudes, and behavior of others. These studies were typically conducted in relatively small social circles. Fast track to modern day, the rise of online social networks (OSNs) has seen a rapid expansion in social network size and the role of opinion leaders has become increasingly crucial in shaping public opinion and driving online discourse. It is widely recognized that developing algorithms that can detect opinion leaders is crucial. There are a range of application areas in business intelligence, social monitoring the spread of (mis)information and mitigating the negative impact on public discourse \cite{bamakan2019opinion}. 

Over the past few decades, empirical research in psychology has explored the phenomenon of opinion evolution during interpersonal interactions. Studies have shown that people tend to modify their opinions to seek similarity with others in the group, highlighting the high interdependence of individual opinions. The combined effects of the influences from cultural norms, mass media and interactions are collectively known as social influence. The concept of opinion leader was first introduced in the hypothesis of \textit{two-step flow of communication} \cite{lazarsfeld1968people}. It posited that the influence from mass media first reaches opinion leaders, who subsequently disseminate it to their followers or associates. 

In recent years, numerous review papers have discussed the related research topics. Riquelme et al. provided an extensive survey on activity, popularity and influence measures that rank influential users in Twitter network \cite{riquelme2016measuring}. Bamakan et al. categorised the characteristics of opinion leaders and the approaches for opinion leader detection \cite{bamakan2019opinion}. A great deal of existing work focus on proxies for opinion-leaders which is to see how information diffuses on the social network statistically, without checking for: (i) whether this information has influence for a topic, and (ii) how is influence actually exerted and by whom. Part of the challenge is the lack of well labeled data sets (need to label topic-specific influence) of sufficient size across diverse topics. Panchendrarajan et al. conducted a comprehensive survey on topic-based influential user detection \cite{panchendrarajan2023topic}. However, there is a lack of consensus between definitions and methods of what constitutes a holistic view of opinion leader across disciplines. In contrast to previous reviews, this review paper focuses on identifying influential nodes in OSNs and providing a cross-disciplinary definition of opinion leaders in relation to social psychology foundational knowledge.

\begin{figure*}[!ht]
    \centering
    \includegraphics[width=0.9\linewidth]{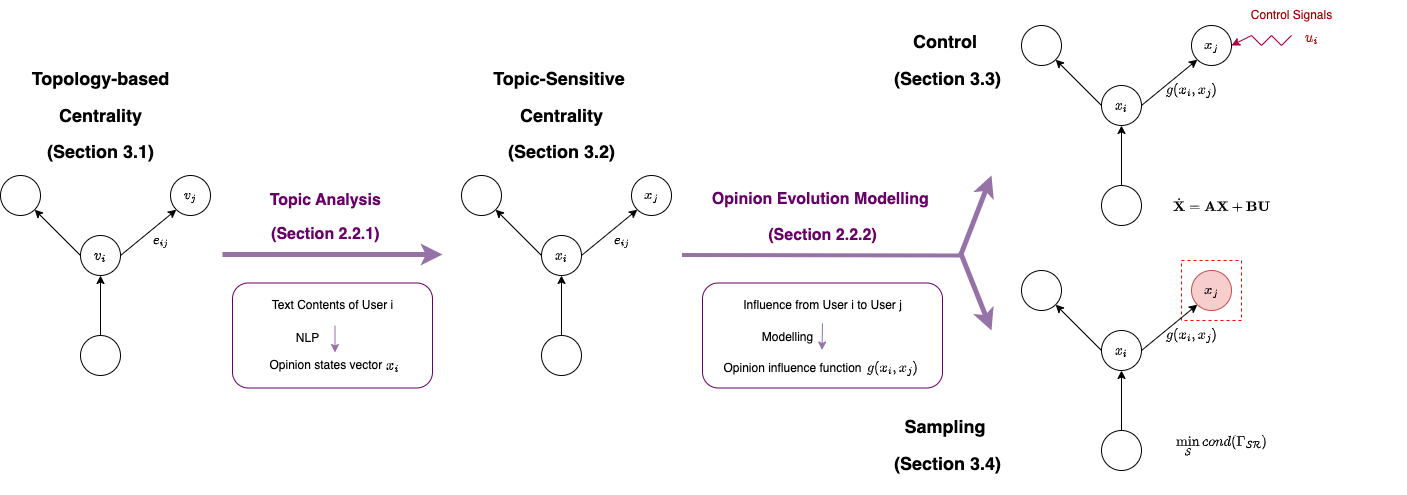}
    \caption{Relationship Among Four Methodologies. The Topology-based Centrality methodologies (3.1) focus on the graphical structure of the network. Real-time content of user $i$ can be represented as an opinion states vector denoted as $x_i$ using Natural Language Processing (NLP). By incorporating Topical Analysis, the Topic-Sensitive Centrality methods (3.2) integrate topical information with the graph structure to identify opinion leaders within specific topics. In opinion evolution modelling, the influence from User $i$ to User $j$ can be modelled as a function $g(x_i,x_j)$. Given the dynamic influence model, the Control methodologies (3.3) aim to find individuals who have the ability to steer the opinion direction by considering the control signals $u$. The Graph Sampling methodologies (3.4) seek to identify specific users who can accurately reconstruct the entire opinion network.}
    \label{meth}
\end{figure*}

In this paper, we categorise the opinion identification methods into four main categories, \textit{Topology-based Centrality}, \textit{Topic-sensitive Centrality}, \textit{Control- and Sampling-based Centrality}. These categories define opinion leaders in distinct ways and ingest different data features. Topology-based centrality mainly concentrates on the network structure. In this context, opinion leaders are defined as individuals who occupy the most significant position within the social group. When user semantic content is taken into consideration, the Topic-Sensitive Centrality facilitates the identification of opinion leaders within specific topics. This approach helps identify influential users capable of disseminating topic-related information and influencing opinions within specific contexts. Additionally, real-time content can be utilised as a representation of the dynamic opinion states of users, which can be used to build a mathematical model to describe the evolution of opinion states. Leveraging the dynamic influence model, control methodologies aim to identify individuals who can steer the direction of overall opinion. Finally, graph sampling methodologies focus on identifying a specific subset of opinion leaders who, despite their limited numbers, can be instrumental in reconstructing the comprehensive opinion network. As illustrated in Figure.\ref{meth}, we show the these four methodologies, and then go on in rest of paper to offer a deeper understanding of each method. 

In table.\ref{table:1}, we provide the notations that are used in this paper. 


\begin{table*}[!ht]
\centering
\resizebox{\linewidth}{!}{\begin{tabular}{ |c l l| }
 \hline
 Variable & Definition & Description \\
 \hline
 $\mathcal{G}$ & A graph of a network & OSN represented as a graph with users as nodes and social connections as edges\\ 
 $\mathcal{V}$ & The set of nodes in the graph & The set of users in the OSN graph\\  
 $\mathcal{E}$ & The set of edges in the graph & The set of connection relationships in the OSN graph\\
 $v_i$ & An arbitrary node (e.g. social account) in the graph $v_i \in \mathcal{V}$ & User $i$ in the OSN\\
 $e_{i,j}$ & An arbitrary link in the graph $e_{i,j} \in \mathcal{E}$ & Social connection(e.g. follow) from User $i$ to User$j$\\
 $N$ & Number of nodes in the graph & The number of users in the OSN graph\\
 $\mathbf{A}$ & The adjacency matrix of the graph & Matrix representation of the OSN network\\
 $a_{i,j}$ & The $(i,j)$th entry of the matrix $\mathbf{A}$ & The presence or absence state of a connection from user $i$ to user $j$\\
 $x_{i}(t)$ & The opinion state of user $i$ at time t & Content posted by user $i$ at time $t$ transformed into numerical value $x_{i}(t)$\\
 $x_i$ & The opinion states vector of user i & Collection of opinions $x_i(t)$ integrated into a vector $x_i$\\
 $X$ & The opinion states matrix of all users & Each column represents the opinion states vector of a user\\
 $f(\cdot)$ & The self opinion dynamic function & It describes self opinion dynamic over time\\
 $g(\cdot)$ & The opinion influence function & It describes the influence between connected nodes\\
 $\textbf{u}(t)$ & The control signals at time $t$ & Input signals imposed on nodes to control the system.   \\
 $\textbf{B}$ & An input matrix & It identifies the nodes that are controlled by the input vector $\textbf{u}(t)$\\
 $\textbf{C}$ & The controllability matrix & The dimension of the controllable subspace of the system is given by the rank of $\textbf{C}$ \\
 $C(i)$ & The exact control centrality of node $i$ & The dimension of controllable subspace of node $i$ with exact values in $\textbf{A}$ and $\textbf{B}$\\
 $C_g(i)$ & The structure control centrality of node $i$ & The maximum dimension of controllable subspace of node $i$ for varying free parameters. \\
 $\bm{\Gamma}$ & The orthogonal subspace of graph signal & Determine the orthogonal subspace of time-varying graph signal.\\
 $\mathcal{R}$ & The graph frequency set & Represent the bandwidth of the networked dynamics to the orthogonal subspace. \\
 $\mathcal{S}$ & The sampling node set & Critical node set that ensure the complete recovery of networked dynamics.\\
\hline
\end{tabular}}
\caption{Notations Used In This Paper.}
\label{table:1}
\end{table*}

\section{Background}

\subsection{Opinion Leader Definition in Social Science}

The concept of an 'opinion leader' in online social media, originally derived from social theory, plays a pivotal role in understanding the dynamics of digital communication networks. This notion aligns closely with the node influence metrics from graph theory. Such applications are crucial for gauging the influence of individuals in social networks and identifying key infrastructure nodes in transportation networks. However, since the early 2000s, social scientists have raised concerns about the adequacy of centrality indices in fully capturing the nuances of node influence in these contexts. 

To clarify the distinctions between graph theory and social theory in defining ‘opinion leader’, we present an overview of the evolution of this concept within social theory. We categorize the definition of opinion leaders based on three main characteristics: spread,impact,and representativeness. The spread, the foremost attribute, originates from the theory of “the two-step flow of communication” \cite{lazarsfeld1968people}\cite{katz1957two}, highlighting that information from mass media first reach opinion leaders who then disseminate it to less active segments. The impact dimension is also underscored in various studies, portraying opinion leaders as individuals capable of  affecting others’ opinions, attitudes and behaviors in an appropriate way \cite{hellevik1991opinion} \cite{rogers1962methods}. Representative, as the third characteristic, positions opinion leaders as trusted and influential within their groups, reflecting collective viewpoints\cite{corey1971people}.

In the context of online social media, the role of opinion leaders has evolved beyond traditional communication models. Presently, they are recognized as individuals who significantly influence others' opinions through online interactions. These leaders are pivotal in various sectors, including marketing, political science, and public health, effectively shaping public opinion \cite{bamakan2019opinion}.

\subsection{Technical Background}

\subsubsection{Topic Analysis}
Topic analysis involves the utilization of natural language processing (NLP) to detect the topic-related semantic structures from human language. In this paper, we mainly employ two types of topic analysis: topic modelling and opinion representation \cite{10.1145/3297662.3365819}. Topic modelling utilizes statistical modelling approaches to assign topic probability distributions to user-generated content. On the other hand, opinion representation is a task of classifying the content into opinion state vectors associated with specific topics.

\subsubsection{Opinion Evolution Modelling}
As the effect of “word-of-mouth”, people are likely to be influenced by the idea of their friends in the process of agricultural innovation, adoption of medical, and new product promotion. To explain how individuals develop their opinions towards various topics over time, a formal model of the opinion evolution in a group was proposed by French in 1956\cite{french1956formal}. In French's formal theory, the discrepancy of opinions $x_i^t$ and $x_j^t$ determines the effect from influencer $j$ to recipient $i$. So the influence effect is determined to be proportional to the size of the difference between their opinions $g(x_j^t, x_i^t) = (x_j^t-x_i^t)$. Beyond the function, there may include influence weights ($w_{ij}$) representing the strength of the effect. Formally, social pressure on the recipient $i$ is the sum of the effect from all influencers j conditioned by the weight ($w_{ij}$) of the tie between i and j. The self-weight ($w_{ii}$) of the recipient $i$ represent to what degree the recipient is anchored on his previous position $(-1.0 \leq w_{ii} \leq 1.0)$ \cite{myers1982polarizing}. The influence process takes place gradually, as the influencer changes its position over time and influences the recipient toward its position. For each recipient, the discrete-time interpersonal influence mechanism can be describe as a ordinary differential equation

\begin{equation} \label{funi}
x_i^{t+1} = w_{ii}x_i^{t} +\sum_{j, j\neq i}w_{ij}(x_j^t-x_i^t)
\end{equation}

From there, social influence models were developed to explain social phenomena such as opinion clustering or controversy. To capture the complexity of opinion evolution, researchers have considered both linear and non-linear models. One example of a linear model is the French-DeGroot model, which introduced a more general form using Markov Chain processes to illustrate how social influence leads to opinion consensus \cite{degroot1974reaching}. However, opinion consensus is not the only outcome from group discussions \cite{R5}. Non-linear models, such as the Hegselmann-Krause model, have been proposed to incorporate a bounded confidence attribute that limits the influence of opposing opinions\cite{hegselmann2002opinion}.

In general, the following equation (\ref{eq:opinion}) represents broader dynamic in both linear and Non-linear models. Here, $\dot{x}(i)$ denotes the rate of change of opinions for agent i, $A$ symbolizes the graph structure, $f_i(x_i)$ represents the self dynamic, $g(x_i,x_j)$ indicates the influence from agent j to agent i, and $u_i$ represents the external social signal influencing agent $i$.

\begin{equation}\label{eq:opinion}
    \dot{x}_i=f_i(x_i)+\sum^n_{j /neq i}g(x_i,x_j).
\end{equation}



\section{Methodology}

The definition of being influential point is ambiguous, leading to the development of various measures for identifying opinion leaders. Here, we categorise detection methods into four groups: topology-based centrality, topic-sensitive centrality, control and graph sampling. 

Centrality is the most commonly employed method, operating under the assumption that opinion leaders are structurally important nodes within a social network. We introduce three traditional measures - Degree, Closeness and Betweenness - and explore their application in the context of online social network. Beyond considering the individual status of neighbours, we delve into a group of eigenvector-based centralities such as Eigenvector, Katz, PageRank. PageRank, for instance, is widely used in topic-sensitive centralities to fit various social network assumptions. In addition to these measures, we also introduce two dynamical measures - Maximization and SIR - which account for the dynamic states of nodes. 

Recognising that the influence of these leaders can fluctuate based on various topic field, the topic-sensitive centrality approaches incorporate topical attributes into the analysis. For instance, topic analysis can be employed to determine the novelty and similarity of content on OSN, which lead to InfluenceRank, TwitterRank, TopicSimilarRank, and ClusterRank. Simultaneously, the dynamics of opinion are analysed based on a particular topic, leading to the creation of OpinionRank, Dynamic OpinionRank, TrustRank and InfluenceModellingRank.

Given the opinion dynamic modelling, social influence can be analysed via control and sampling methodologies. Control methodologies identify influential nodes by assessing their ability to influence the states of others in the network.

Another approach is graph sampling, which determines the most influential nodes by studying whether the samples on these nodes can ensure the complete recovery of the whole networked dynamics. This is generally achieved by determining an orthogonal subspace of the networked opinion vector, and then evaluating the importance mapping from the orthogonal basis to the node set. This section reviews how to build the orthogonal subspace from the spatial and spatial \& temporal dynamics correlation, and the graph sampling-based ranking strategies. 


\subsection{Topology-based Centrality}
Centrality in graph theory and network analysis is a fundamental concept that refers to the importance of a node within a network. In the context of social network, centrality measures help identify users who have extensive connections with other members of a network. 
Bavelas first introduce the idea of centrality to human communication network, aiming to explain the influence in group processes\cite{bavelas1948mathematical}. Bavelas proposed that an individual strategically positioned on the shortest communication path connecting pairs of others within a group occupies a central position. Subsequently, various methods for detecting opinion leader based on centrality have been proposed. 


\subsubsection{Degree Centrality}
Degree centrality is the number of connections a node has in a network. Freeman presented Degree Centrality in social network, which is rooted in the belief that an individual’s significance within a group is tied to the number of people they are connected to or interact with \cite{freeman1979centrality}. In real-world case, the node with the highest degree is the user that directly interacts with many other users within the network. This method is intuitive to the definition of influence, whereas the global structural of the graph is not considered.

\subsubsection{Betweenness Centrality}
Betweenness centrality is based on the number of times a node lies on the shortest path between two other nodes in the network\cite{freeman1979centrality}. In online social network, user with high Betweenness centrality operates like a bridge in the shortest paths between possible user pairs. Closeness and Betweenness centrality are challenging to apply in large-scale networks, and have been proved to be unstable in some cross-sectional and temporal networks.

\subsubsection{Closeness Centrality}
Taking consideration of indirect link using the path length, the closeness centrality extends the local centrality to global centrality. The basic idea of closeness centrality is that the node with high closeness centrality can spread the information to other nodes quickly. In this case, the position of one point in the network is more essential than the number of links it own. In online social network, users with high closeness centrality have been proved to be effective spreaders of information by measuring the diffusion effect \cite{yang2018identifying}. However, Closeness centrality is very sensitive to a large distance or missing link due to considering the distance of each pair.

\subsubsection{Eigenvector Centrality}
Eigenvector centrality of a node is calculated as the weighted sum of the centralities of its neighbors, with the weights determined by the strength of the connections between the node and its neighbors. Therefore, this measure can be used to measure the level of influence of each node, where the higher score the greater level of influence. Eigenvector centrality is designed to differ from the former measures when the network contains high-degree nodes connected to many low-degree nodes or low-degree nodes connected to a few high-degree nodes. The disadvantage of Eigenvector Centrality is that it has limitations when applied to directed networks. A node can receive a score of zero in the absence of incoming links, resulting in no contribution to the centrality metric of other nodes.

\subsubsection{Katz Centrality}

Katz and PageRank are variants of the eigenvector centrality. Katz centrality takes into account both the number of direct connections a node has and the connections of its neighbours, which can be less sensitive to the size of the network and proved stable ranking\cite{katz1953new}. The limitation of Katz centrality is that it can be influenced by new links to a particular group of nodes. 

\subsubsection{PageRank Centrality}
To mitigate the impact of spendthrift nodes on centrality scores, PageRank reduce the weight of ingoing links from these nodes. In PageRank, the weight of an incoming link is proportional to the PageRank score of the node it originates from\cite{page1999pagerank}. Compared to Katz centrality, PageRank add the scaling factor which gives it the ability to penalise nodes that are linked to from many low-quality nodes and reward nodes that are linked to from high-quality nodes. In this way, the PageRank centrality mitigates the impact of nodes with many outgoing links, and instead focuses on the quality of the incoming links, rather than the quantity.

\subsubsection{HITS}
As PageRank, Hyperlink-Induced Topic Search(HITS) is also a link-based ranking algorithm to determine the importance of the node\cite{kleinberg1999authoritative}. The intuition of HITS is that Authority score and Hub score are both allocated to each web page. Assuming that high-quality Hub usually point to high-quality Authorities, and high-quality Authority is pointed by high-quality Hubs. As a result, the Authority score is proportional to the total hub scores of the Hubs that link to it. In online social network, the algorithm search the influential accounts by collecting the query-related accounts and then ranking only by the network structure instead of textual contents.  

\subsubsection{SPEAR}
Yeung et al. introduced the terms experts and expertise for resource discovery\cite{yeung2011spear}. Assuming that a user’s expertise depends on the quality of the resources they have collected and the quality of resources is depend on the expertise of other users who have assigned relevant tags, Spamming-Resistant Expertise Analysis and Ranking(SPEAR) was introduced to rank users in online knowledge communities. SPEAR is a graph-based algorithm similar to the HITS algorithm implementing the concept of expertise. Later in 2016, Shinde and Girase proposed the modified SPEAR algorithm\cite{shinde2016identification} where the expertise of user is based on different topics. In the topic-specific SPEAR algorithm, the credit score function considers not only time, but also number of comments, number of likes, word count and all.

\subsubsection{TunkRank}
Tunkelang introduced TunkRank, a measure of user influence based on PageRank\cite{tunkelang2009a}. TunkRank operates on three assumptions: 1)influence power of a certain influencer corresponds to expected number of audiences who read a tweet from the influencer, 2) the probability of audience reading a tweet based on the number of accounts they follow, and 3) the audience has a constant probability to retweet the seen tweet. The expected number of people who read the tweet can be recursively calculated based on the equally distributed probability of each follower read the tweet and the constant probability that user will retweet the tweet.

\subsubsection{Dynamical Influence}
Dynamical influence is a centrality measure that takes into account the interplay between network structure and the dynamical state of nodes. This is a departure from classical centrality measures which rely solely on topology. In the context of social network, the dynamical influence process can be used to explain the dynamics of idea adoption. In this scenario, opinion leaders are defined as the key individuals who can trigger a significant cascade of influence. The challenge lies in identifying these key individuals, which is essentially an influence optimization problem. The goal is to target an initial set of nodes with the greatest influence spread, thereby promoting information to a large fraction of the network. The maximization of information flow was first considered as a discrete optimization problem by Kempe et al.\cite{kempe2003maximizing}. They discussed models for how influence propagates through online social networks, and proposed a greedy hill climbing approach of identifying the most influential nodes which provide provable approximation guarantees. Zhao\cite{zhao2015seismic} built a statistical model SEISMIC building on the theory of self-exciting point processes to model the information cascades. SEISMIC provides an extensible framework for predicting information cascades. It requires no feature engineering and can scaling linearly with the number of observed reshapes.

\subsubsection{SIR}
The Susceptible-Infected-Recovered(SIR) model is another algorithm that considers the dynamical state of nodes. The SIR mathematical model was originally designed to describe the spread of infectious diseases in a population. The model divides the dynamical state of population into three categories: Susceptible(S), Infected(I), and Recovered(R). The SIR model has also been used to model the spread of information in a network, where nodes can be thought of as either susceptible to influence, influenced, or recovered from the influence. When applied to identify opinion leaders, the opinion leaders are set as the initially infected nodes, and the probability of an infection depends on the influence from the opinion leader.

\subsubsection{Meta-Centrality and Learning Meta-Ranks}
There are a multitudes of other meta-centrality approaches such as Centripetal Centrality, combining multiple centrality approaches \cite{R2}. Such functional combination approaches open up the more reasonable method of using deep learning to learn new centrality measures \cite{R3}. However, recent work recognises that understanding the topic context is important to not only directing centrality measures to be more precise, but also incorporating knowledge of influence behaviour into the centrality measures \cite{R4}.

\subsection{Topic-sensitive Centrality}
Opinion leaders are identified based on various characteristics that align with diverse social groups. Analysis of dynamic influence across topics and time has demonstrated that ordinary users can gain influence by focusing on a single topic\cite{cha2010measuring}. Recognising that the influence of these leaders can fluctuate based on various topic field, the topic-sensitive centrality approaches incorporate topical attributes into the analysis. Several topic-sensitive ranking methods have been developed to determine the topical influence of users and their capacity to disseminate information or influence opinion on specific topics. 
Simultaneously, the dynamics of opinion can be analysed based on a particular topic.

\subsubsection{InfluenceRank}
Topical analysis can be used to quantify the novelty of certain content by representing each content as a document and reducing the dimensionality using Latent Dirichlet Allocation (LDA). The InfluenceRank algorithm uses the topical analysis to measure the importance and novelty of a blog in comparison to other blogs\cite{song2007identifying}. With the feature vectors that represent the topic distribution, the dissimilarity can be calculated using cosine similarity. InfluenceRank outperforms other algorithms in terms of coverage, diversity and distortion.

\subsubsection{TwitterRank}
TwitterRank is a variant of the TunkRank algorithm that incorporates topical similarity in the calculation of influence. The phenomenon of “homophily” has been observed in various network ties, including information transfer, friendship, and marriage\cite{mcpherson2001birds}. Weng et al. demonstrated that “homophily” also exists in the context of Twitter, where users tend to follow those who share similar topical interest\cite{weng2010twitterrank}. TwitterRank was proposed based on this finding, measuring influence by considering both topical similarity and link structure. However, users’ topical interests can change over time, and as a result, the freshness of their activities needs to be taken into account. Dhali et al.\cite{dhali2020attribute} addressed this issue by proposing TemporalTwitterRank, a modified algorithm that estimates transition probabilities using topic profile vectors. By emphasizing the temporal dimension of users' activities, TemporalTwitterRank provides a more comprehensive assessment of influence.

\subsubsection{TopicSimilarRank}
Wang et al. proposed the TopicSimilarRank algorithm considering the user’s own influence and difference in influential values caused by responses from others. The TopicSimilarRank algorithm is inspired by TwitterRank and takes into account topic similarity, user attributes, interactive information, and network structure. To construct the weighted network, the users can be seen as a set of weighted nodes, and the reposts and comments can be seen as edges with weights represented by similarity values between users. Then the directed and weighted graph can reflect the influential relationships between users. The experiment analysis indicates that TopicSimilarRank is well-suited for mining opinion leaders in topic domains. Similarly, Eliacit et al. \cite{eliacik2018influential} developed three metrics - User Trust (degree of friendship, expertise and activity), Influence Period and Similarity - to construct a weighted influence network. Influence rank was calculated based on the PageRank Algorithm. The empirical experiment demonstrates that considering the ranking of users enhances the accuracy of sentiment classification in the community.

\subsubsection{ClusterRank}
To identify the most influential authors for a specific topic, Pal  and Counts proposed a set of features, including both nodal and topical metrics, to describe the authors in various topic fields\cite{pal2011identifying}. To reflect the impact of users with respect to one topic, various features are selected for original tweets, conversational tweets and repeat tweets. ClusterRank process includes using probabilistic clustering on this feature space, within-cluster ranking procedure and producing a list of top authors for a given topic. The experiment showed that topical signal and mention impact are two critical features to determine the ranking.

\subsubsection{OpinionRank}
OpinionRank considers both the dynamic of information influence and the dynamic of forming opinions. In 2009, Zhou and Zeng introduced the concept of opinion networks and OpinionRank algorithm to rank the nodes based on their opinion scores\cite{zhou2009finding}. In this context, a weighted link in the opinion networks represents the opinion orientation from opinion sender to opinion receiver. For instance, in a review website, the opinion receiver is the original review writer and the opinion sender is the comment writer under the review. The opinion orientation can be calculated as the average opinion score after assigning an opinion score to each word. Experimental results have indicated that sentiment factors significantly influence social network analysis.

\subsubsection{Maximization}
Huang et al. introduced the Positive Opinion Leader Detection (POLD) to track the public formation process\cite{huang2014finding}. POLD constructs multiple opinion networks on comment networks rather than user networks. The comment network takes into account the time interval between comments, assuming that influence weakens with increasing intervals. Applying POLD to the comments of news reveals that the most influential comments and users vary over time. Dong et al. further hypothesised that influence only occurs when a recipient posts within a certain time interval after the influencer\cite{dong2019mining}. The weight of edges in this network is modelled based on the time gap between the posts by influencer and recipient. 

\subsubsection{TrustRank}

Chen et al. proposed the TrustRank considering both positive and negative opinions\cite{chen2014identifying}. TrustRank constructs a network with direct and indirect sentiment labelled links. The construction has 4 phases: 1) set up a basic network, 2) label the links, 3) infer the sign, and 4) transform the post network to user network. During the construction of network, both explicit link and implicit link are considered. The explicit link is denoted by reply and citation, and implicit link infers the semantic similarities between posts. TrustRank outperforms other PageRank-like models on the online comments of a real forum.

\subsubsection{InfluenceModellingRank}
In our previous work, we proposed a method to model the evolution of personal opinions as an ordinary differential equation (ODE)\cite{10068693}. To account for the influence of influencers on one recipient's opinion, we employed French's formal theory\cite{french1956formal} to model the social influence effect. This effect is determined by the discrepancy of their opinions and the influence weight representing the strength of the effect. To compute the influence weight, we utilized a collection of following links and posts. By assigning the influence weights as link weights and using the PageRank algorithm, we were able to rank the users based on their influence weight. The resulting \textit{InfluenceModellingRank} provides a metric for understanding the opinion influence dynamics in social networks.

\subsection{Influence based on Control Centrality}
Social Influence is roughly defined as follows: Given two individuals $u$,$v$ in a social network, $u$ exerts the power on $v$, that is, $u$ has the effect of changing the opinion of $v$ in a direct or indirect way \cite{cercel2014opinion}. The influence of an individual in the social network is affected by the self-dynamics of the individual's behavior, coupling dynamics between individuals, and the network structure of the social network. Metrics for influence based on the previous centrality measures mainly consider the network topology of the social network. When we consider both the social network structure and the dynamics of each node, it is natural for us to ask the following questions: 
\begin{enumerate}
    \item whether it is possible for a node to influence other nodes to any desired state  
    \item how many nodes' states can be influenced by one nodes
Therefore, it is reasonable to introduce controllability in complex network to quantify the influence of each node and detect the influential node.
\end{enumerate}
Here we introduce the concept of controllability in complex networks to identify influential nodes. The analysis framework we introduce here to identify influence nodes can be generally applied in social networks, which reflects in following perspectives: 1) this framework can be used in any linear dynamics and does not need to know the specific dynamic functions; 2) only the network topology of the social network is needed, and even the weights of connections are not necessary to know.

\subsubsection{Kalman's criterion of controllability}

Consider a complex system described by a directed weighted network of $N$ nodes, the dynamics of a linear time-invariant (LTI) system can be described as 
\begin{equation}
    \dot{\textbf{x}}(t)=\textbf{Ax}(t)+\textbf{Bu}(t),
\end{equation}
where $\textbf{x}(t)=(x_1(t),x_2(t), \cdots x_N(t))^{\top} \in \mathbb{R}^N$ captures the state of each node at time $t$. $\textbf{A}\in \mathbb{R}^{N \times N}$ is an $N \times N$ matrix describing the weighted connection of the network. The matrix element $a_{ij} \in \mathbb{R}$ gives the strength that node $j$ affects node $i$. $\textbf{B}\in \mathbb{R}^{N \times M}$ is an $N \times M$ input matrix ($M \le N$) identifying the nodes that are controlled by the time-dependent input vector $\textbf{u}(t) = (u_1(t), u_2(t), \cdots, u_M(t)) \in \mathbb{R}^{M}$ with $M$ independent signals imposed by the controller. The matrix element $b_{ij} \in \mathbb{R}$ represents the coupling strength between the input signal $u_j(t)$ and node $i$. The controllability of the LTI system can be checked by the best known Kalman's  rank condition \cite{kalman1963mathematical} which states that the LTI system is controllable if and only if the $N \times NM$ controllability matrix
\begin{equation}
    \textbf{C} \equiv [\textbf{B}, \textbf{AB}, \textbf{A}^2\textbf{B},\cdots,\textbf{A}^{N-1}\textbf{B}]
\end{equation}
has full rank, i.e.,
\begin{equation}
    {\rm{rank}} ~\textbf{C} = N.
\end{equation}
When the system $(\textbf{A}, \textbf{B})$ is not controllable, the dimension of the controllable subspace is ${\rm rank}~\textbf{C}$, where ${\rm rank}~\textbf{C}<N$.

\subsubsection{Exact control centrality}
One thing we are interested in is how many dimensions of the subspace of the system can be controlled by a single node. Here, we use ${\rm rank}~\textbf{C}^{(i)}$ to capture the ability of $i$ in controlling other nodes in the networked system.  Mathematically, ${\rm rank}~\textbf{C}^{(i)}$ captures the dimension of the controllable subspace or the size of the controllable subsystem when we only control node $i$. The exact control centrality of node $i$ is defined as 
\begin{equation}
    C(i) \equiv {\rm rank}~(\textbf{C}^{(i)}),
\end{equation}
where the $\textbf{B}$ in matrix $\textbf{C}$ reduces to the vector $\textbf{b}^i$ with a single nonzero entry, e.g.  $\textbf{b}^i=[0, 0, \cdots, b_i, \cdots]^{\top}$. By calculating the exact control centrality of each node in the networked system, the most powerful nodes in controlling the whole networked system can be identified. In a social network, with exact parameters, users with higher $C(i)$ can affect more users' opinions. Therefore, we can find the most influential nodes according to exact control centrality. 

\subsubsection{Structural controllability}
When we know the exact network structure and the weight of each connection in the social network, the influence of each user can be ranked by the exact control centrality. However, there exist some limitations when the exact control centrality method is applied to analyze the influence of each user. $C(i)$ is sensitive to the perturbations of elements of matrix $\textbf{C}^{(i)}$, especially in a large matrix. Usually, the social network contains a large number of users, so estimating the influence of each user by the exact control centrality has a high requirement of the accuracy of the weight of connections. The second limitation is that in social networks, the system parameters are not precisely known, e.g. the elements in matrix $\textbf{A}$ are not exactly known. We only know whether there exists an influence between two users but are not able to measure the weights of the influence between them. Hence, it is difficult to numerically verify Kalman's controllability rank condition using fixed weights. To solve this problem, the concept of structural control \cite{lin1974structural} can be introduced to measure the influence in social networks. The power of structural controllability comes from the fact that if a system is controllable the it is controllable for almost all possible paramter realizations \cite{liu2016control}.

An LTI system $(\textbf{A}, \textbf{B})$ is a structured system if the elements in $\textbf{A}$ and $\textbf{B}$ are either fixed zeros or independent free parameters. Apparently, ${\rm rank}~(\textbf{C})$ varies as a function of the free parameters of $\textbf{A}$ and $\textbf{B}$. It achieves the maximal value for all but an exceptional set of values of the free parameters. This maximal value is called the generic rank of the controllability matrix $\textbf{C}$, denoted as ${\rm rank_g}(\textbf{C})$, which represents the generic dimension of the controllable subspace. The system $(\textbf{A}, \textbf{B})$ is structurally controllable if we can set the nonzero elements in $\textbf{A}$ and $\textbf{B}$ such that the resulting system satisfies ${\rm rank_{g}}~\textbf{C} =N$. The minimum number of nodes which control the state of the full system can be identified by mapping this problem to a pure graph-theoretical problem called maximum matching \cite{murota2010matrices, liu2011controllability}. In social influence networks, the subset of these nodes are the most influential nodes which can influence the state of all nodes in the network.

\subsubsection{Structural control centrality}
Correspondingly, to measure the dimension of the controllable space of one node without the exact information of system parameters, the concept of structural control centrality has been introduced, which can be defined as \cite{liu2012control}
\begin{equation}
    C_g(i) \equiv {\rm rank_g}~(\textbf{C}^{(i)}).
\end{equation}
The structure control centrality is an upper bound of exact control centrality for all admissible numerical realizations of the controllable matrix $\textbf{C}$. So, in an influence network, the structure control centrality is a method to estimate the largest number of users that can be affected by one user with clearly known connections between users. To calculate $C_g(i)$, we need to introduce some concepts in graph theory. A node $j$ is called accessible if there always exists at least one directed path from the input nodes to $j$. A stem is a directed path starting from an input node, so that no nodes appear more than once in it. $C_g(i)$ can be calculated according to Hosoe's controllable subspace theorem \cite{hosoe1980determination}:
\begin{equation}
    {\rm rank}_g(\textbf{C}) = \mathop{max}\limits_{G_s \in G}|E(G_s)|,
\end{equation}
where $G_s$ is the subgraph of the accessible part of $G$ only consists of stems and cycles and $|E(G_s)|$ represents the number of edges in $G_s$. The action space may take many forms from inserting control signals to rewiring the graph structure \cite{10552434}, this falls outside this review.

\subsection{Graph Sampling \& Recovery}
Another idea to determine the most influential nodes over a network leverages whether these nodes can be sampled to recover the whole networked dynamics. This refers to as graph signal sampling and recovery techniques, which aim to compress the time series of high-dimensional and dependent networked dynamics via a subset of critical nodes, whose dynamics can guarantee the recovery of the whole networked data. From the theoretical perspective, this includes low-rank matrix completion, optimization with Laplacian constraints, the spatial, and the temporal dependency analysis, whereby the former studies the correlation or hidden high-dimensional dependency among the set of nodes, and the latter focuses on the events at which time steps would be the trigger or with higher importance.

\subsubsection{Low-Rank Approximation}
Low-rank matrix completion aims to recover the whole matrix from the known entries (samples) \cite{8233154,8292837,donavalli2012low}. The typical approach is to minimize the rank of the recovered matrix by the singular values, constrained by the values of the samples \cite{candes2012exact}. CUR \cite{mahoney2009cur} serves another popular method family, leveraging the sampled rows and columns to recover the whole data matrix. 
In the context of opinion leader identification, the rows of the matrix are the time-serial language embedding of different users, while the columns represent the sampling time indices.  In this view, the aim is to find the minimum set of users whose embedding can recover the whole data matrix. This set of users constructs a low-rank core of the data matrix, which contributes significantly to the whole information entropy of the data. By capturing their tendency of the topics, the information flow of all the users can be approximately obtained. 

\subsubsection{Optimization with Laplacian Penalty}
The eigenvalues and eigenvectors of the Laplacian matrix of the graph structure serve as the graph frequency domain. Under the assumption that critical users capture a large portion of the information entropy of the networked dynamic data, the Laplacian penalty \cite{7979523} can be set to constrain the graph bandwidth of the data for critical user identification. In this view, the critical users found by this approach represent the approximated data with minimum graph bandwidth concerning the graph structure-based Laplacian matrix, which, however, does not involve the specific dynamic patterns (e.g., ODE or PDE) that govern the evolution of the language embedding propagation \cite{9234614}. As such, the selected users can better represent the graph structure from the frequency domain (other than the topological node domain as stated in Section 2.1), but lack the propagation information for different topic-sensitive patterns.

\subsubsection{Spatial Correlation Analysis}
Spatial correlation analysis tries to determine an orthogonal signal subspace (matrix), e.g., the operational matrix in compressed sensing (CS), or the graph Fourier transform (GFT) operator  \cite{8839864,8865055,9071735}. Then, leveraging the orthogonal subspace, the highly correlated networked data can be compressed by the linear combinations of the subset of the orthogonal bases, which can be mapped to the critical nodes for sampling and recovery purposes.

Compared to the Laplacian penalty strategy, this approach takes into account both the graph structure and the dynamic pattern directly extracted from the data, and is therefore better to provide the critical opinion leader set for different topic and network -sensitive topics. One drawback is the overlook of temporal correlations between different time stamps, which will underestimate the appearance of opinions in the evolution process.

\subsubsection{Spatial \& Temporal Dependency Analysis}
Spatial and temporal dependency analysis aims to determine the critical nodes by considering both the node level and temporal level correlations \cite{9234614}. By combining the temporal correlation information, a more compact dynamic subspace can be derived, which gives rise to a reduction in the number of sampling nodes, and leads to a node importance rank. 

The derivation of the dynamic subspace contains the model-based and data-driven approaches. The model-based approach is to generate an orthogonal subspace leveraging the linearized dynamic model, e.g., via dynamic mode decomposition (DMD) or extended DMD (E-DMD) \cite{dey2022dynamic,brunton2019notes}. Such a model-based subspace compresses the networked dynamics via the spatial and temporal correlations. The opinion leader identification is then converted to the sampling and recovering problem that selects the critical nodes to make truncated subspace full column rank. 

When the model is unavailable (e.g., difficult to pursue a linear regression), the data-driven methods are well-suited to derive the dynamic subspace. To be specific, by pursuing a compact singular value decomposition of the data, the dynamic subspace can be constructed by the Kronecker product of the left and right singular vectors with non-zero singular values. After the derivation of the dynamic subspace, the important users can be derived by the greedy selection of rows to maximize the least singular value of the subspace.

\subsection{Evaluation Method}
The evaluation of Opinion Leader Detection methods is not straightforward, and various papers use different evaluation methods. Unfortunately, there is no agreement on which evaluation method is the best. Nonetheless, some evaluation methods are still commonly used and will be discussed in this section.

\subsubsection{Descriptive Methods}
In social sciences, descriptive methods are often utilized to identify opinion leaders. One of the most popular descriptive methods is using experts to rank the opinion leaders in one group network. In this approach, an expert is asked subjectively to rate the comments from users having either a strong or weak influence. The ratings of comments are then combined to determine the influential rate of each user. However, descriptive methods require creating questionnaires and conducting interviews, which are costly and challenging to implement. These descriptive measures have been criticized because they do not consider the role of ordinary users in the information flow process\cite{valente2007identifying}.

\subsubsection{OSN Metrics}
For OSN platforms, the number of followers is the most commonly used metric to determine a user's influence. This approach assumes that each tweet by a user is read by all of their followers. Other metrics such as likes, shares, or mentions are also used to measure user engagement and influence\cite{bruns2014metrics}. On Twitter, these public metrics are accessible through the Twitter Application Programming Interface (API), which is built on communication data and metadata.


\subsubsection{Kendall's $\tau$}

Kendall’s $\tau$ is a statistical measure to determine the similarity between the ranking orders of two variables, regardless of their magnitudes. Kendall’s $\tau$ coefficient ranges from $-1$ to $1$, with a value of -1 indicating complete disagreement between the rankings, and 1 indicating perfect agreement between the rankings. Kendall’s $Tau$ correlation method is often used in social science research, including the opinion leader detection task, to assess the degree of agreement or disagreement between two rankings.

\section{Case Study}

This section presents a case study of the application of different ranking methods to 2 Twitter datasets widely used by the research community: (1) COVID-19, and (2) feminism debate. In order to gather and prepare data for our pilot study, we relied on the methodology outlined in our previous study \cite{10068693}, which is we identified topic specific active users who posted a minimum level of topic-specific tweets over a specified period. Based on our process, we identified 98 active users for the COVID-19 topic and 180 active users for the feminism debate topic. We then crawled and analyzed 85,946 COVID-19 related tweets and 69,088 feminism related tweets. To pre-process the tweets and capture the vibration of opinion, we used compressed word-embedding vectors \cite{10068693}. For the validation of the different centrality rankings, we selected four topic-filtered matrix rankings: Retweet, Reply, Like, and Quote. These filtered matrix rankings were calculated by considering only the topic-related tweet matrix posted by the group of users.

\subsection{Proof of Concept}

We first computed the previously reviewed three centrality rankings commonly used directly or as part of meta-centrality: Betweenness, Eigenvector, PageRank, as well as one topic-sensitive ranking: InfluenceModel, one control theory ranking, and two previously reviewed compressive graph sampling rankings: MGFT, DGFT. 

The Kendall $\tau$ correlation results are illustrated in Table \ref{table:covid3} and Table\ref{table:feminism3}. Since the absolute value of all Kendall $\tau$ are lower than 0.3 and most of them are close to 0 which means these ranking vectors are likely to be independent.

Our observations backed by logic and topic-sensitive evidence are as follows:
\begin{itemize}
       \item The three classic centrality rankings and topic-sensitive ranking exhibit greater performance similarity with each other. Conversely, the Control ranking and two graph sampling rankings yield distinct results due to their different definitions of influence. This demonstrates that indeed the new definitions offer alternative value.
       \item In Table \ref{table:covid3}, the Control rank exhibits the highest similarity score with Retweet rank, and both GFT rank approaches demonstrates the highest similarity score with Reply rank. This shows key control points is a better indication of retweet relays, whereas graph compression is a better indication of response.  
       \item In Table\ref{table:feminism3}, the MGFT rank achieves the highest similarity scores with all four validation ranks.
\end{itemize}

A general non topic-sensitive horizontal comparison among different ranking strategies is challenging, due to the different criteria utilised by the methods. For instance, in an extreme case where someone that replies to many posts may indicate that they are incorporating and modifying the context, then the graph sampling theory may select it as an influential node, given its potential to contribute to data recovery of all other contexts. Such a user, on the other hand, may not rank highly from a control perspective, meaning they do not control conversation. Then, there may be some overlap that a good representative or control user may have good topographical or topic-sensitive properties, yet their correlation and causality require further studies.

\begin{table*}[!ht]
\centering
\resizebox{\linewidth}{!}{\begin{tabular}{ |c c c c c c c c| }

\hline
$\tau$ Score & Betweenness & Eigenvector & PageRank & InfluenceModel & \textbf{Control} & \textbf{MGFT} & \textbf{DGFT}\\
\hline
\textbf{Retweet} &  0.025&0.055&0.066 &  0.032&\textbf{0.274}&0.046&-0.046\\
\textbf{Reply}   & 0.107&0.03&0.012& 0.084&-0.002&\textbf{0.172}&\textbf{0.206}\\
Like    &  0.125&0.020&0.02&  0.083&0.013&0.156&0.177\\
Quote   & 0.121&0.015&-0.002& 0.080&0.033&0.141&0.161\\

\hline
\end{tabular}}
\caption{Kendall's $\tau$ Score on ordinal categorical COVID-19 dataset.}
\label{table:covid3}
\end{table*}

\begin{table*}[!ht]
\centering
\resizebox{\linewidth}{!}{\begin{tabular}{ |c c c c c c c c| }
\hline
$\tau$ Score & Betweenness & Eigenvector & PageRank & InfluenceModel & Control & \textbf{MGFT} & DGFT\\
\hline

Retweet & 0.041&0.083&0.008& 0.052&0.098&\textbf{0.105}&0.031\\
Reply   &  0.014&0.043&0.010&  0.035&0.108&\textbf{0.116}&0.081\\
Like    &  0.017&0.046&0.01&  0.042&0.104&\textbf{0.122}&0.065\\
Quote   &  0.01&0.082&-0.013&  0.048&0.096&\textbf{0.102}&0.021\\

\hline
\end{tabular}}
\caption{Kendall's $\tau$ Score on ordinal categorical Feminism dataset.}
\label{table:feminism3}
\end{table*}

\subsection{Improvements for Community to Make}

In our current approach, a progressive research flow on ranking/identifying opinion leaders is provided, whereby different ideas leveraging the uses of topology, static topic-sensitive, and opinion differential evolution are reviewed and evaluated. The main challenge would be integrating diverse contents, features and dynamics as input data to find the opinion leaders. Diverse contents refers to incorporation of Cross-platform content analysis—for instance, text content on Twitter, graphical content on Instagram, and video content on TikTok. Various features can include user features, content features and network features. Dynamic nature of social network and human behavior presents another challenge. The continuous flow of data offers opportunities to identify emerging influencers who have the potential to influence,high dimensional data resources.

In current studies, the original content is pre-processed via the word embedding and compression to represent the dynamics of opinions. However, this compression process inevitably lead to information loss, including the delays, the hidden dependency from spatial and temporal perspectives. In future work, to reduce information loss, more complicated opinion representation may be generated to describe the opinion evolution, which will challenge the current raking strategies that leverage the networked dynamics. Consequently, how to design nonlinear sampling and control spaces may be worth studying in the future. Furthermore, it is also noteworthy that even with the sophisticated ODE construction with graph signal evolution and control layer inputs, only the correlation/dependency overlap with the opinion leader set can be identified. In this view, on one-hand, how to build a causality model that represents the causal relations from opinion leaders to the dynamic evolution data requires further studies; on the other hand, the reverse flow from the networked dynamics to infer the opinion leader from the causality perspective remains untouched and is worth studying.

\section{Discussion}

\begin{table*}[h!]
\centering
\resizebox{\linewidth}{!}{\begin{tabular}{ |l l l | }
 \hline
 Psychology Definition & Concept Type & Method\\
 \hline
 Unbiased dynamic based spread in large groups. \cite{lazarsfeld1968people} (Section 3.1) & Spread & Large \textbf{Topology}-based\\
 Individuals who disseminate topic-related information effectively. \cite{katz1957two} (Section 2.2.2) & Receptiveness & \textbf{Topic} Sensitive\\
 Repeat the information to maximize influence based on observed context. \cite{rogers2014diffusion} (Section 3.3) & Control & Graph signal \textbf{control} \\
 Respond to information to capture diverse contexts \cite{corey1971people} (Section 3.4)& Representation & Graph signal \textbf{compression}\\
\hline
\end{tabular}}
\caption{Comparative inter-disciplinary analysis of opinion leader definitions and their detection methodologies}
\label{table:def}
\end{table*}

The purpose of this review is to provide an overview of the development of "opinion leader" concept and detailed comparative analysis of the corresponding detection techniques. The key novelty is to review technical connections and cross-compare different approaches that have origins in: social theory, graph theory, control theory, and graph sampling - with the eventual goal to more holistically describe trusted opinion leaders and their strategies in influence \cite{corey1971people}. Here, we conclude with a comparative analysis of opinion leader definitions and their detection methodologies, as shown in Table \ref{table:def}.

When it comes to topic-sensitive centrality, text content of a user's conversation can be transferred into topic probability distribution or opinion states vector. In the case of the former, topic analysis can be leveraged to determine the novelty and similarity of content. Here, the identified opinion leader exhibits the ability to generate novel content and disseminate topic-related information effectively. In the case of the latter, opinion dynamics are analyzed under a specific topic. The identified opinion leaders in this scenario are users who possess the capacity to influence opinions of others \cite{rogers2014diffusion}. Here, the receptiveness of the recipients is more important than graph structure. Classical models include psychology informed differential equations in small group psychology experiments \cite{french1956formal}\cite{degroot1974reaching}\cite{hegselmann2002opinion}, which we reviewed in Section 2.2.2.

Opinion evolution modeling was developed to explain how individuals develop their opinions towards various topics over time. Researchers have considered both linear and non-linear models to capture the complexity of opinion evolution. With the linear dynamics of opinion modeling, such as the French-DeGroot model, control theory can be broadly applied to dynamic opinion networks, eliminating the need of knowing the specific dynamic functions or the weights of connections. The application of control theory aims to find opinion leaders who can steer the overall direction of public opinion, which we reviewed in Section 3.3.

Graph sampling theory can be applied to identify multiple opinion leaders that minimize redundant information and influence pathways and maximize the overall efficiency of the system in spreading influence. Using either linear model-based and data-driven manners, where the latter does not require the awareness of the dynamic model nor its linearity assumption. By deriving the orthogonal subspace from the data, the data-driven graph sampling method can obtain the opinion leaders who are the representative users in the dynamic opinion networks, which we reviewed in Section 3.4.  

\begin{figure*}[!ht]
    \centering
    \includegraphics[width=0.6\linewidth]{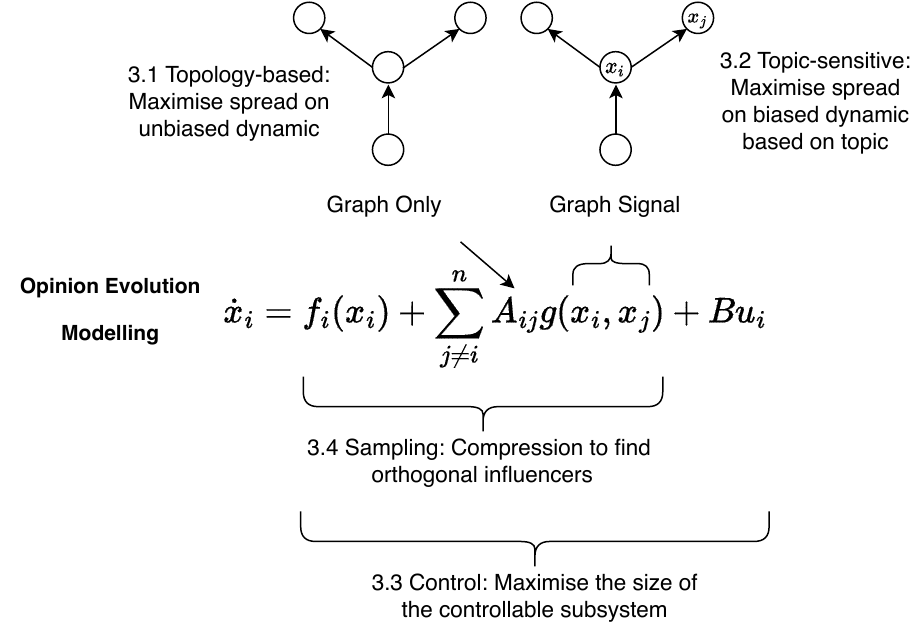}
    \caption{Opinion Evolution Modeling, interlinking four methods: The Opinion Evolution Modeling formula at the center delineates the mathematical model underpinning opinion dynamics. Topology-based detection emphasizes the maximization of spread on an unbiased dynamic represented solely by graph structure $A_{ij}$. Topic-sensitive detection, where the maximization of spread on a biased dynamic is contingent on the topical relevance of the content, signified by the graph signal $x_i$ based on topic. By constructing opinion evolution models as $\dot x_i = f_i(x_i)+\sum_{j\neq i}^n A_{ij}g(x_i,x_j)$, graph sampling methods identify orthogonal influencers to reduce redundancy in message dissemination across the network. Control theory approaches aim to maximize the size of the controllable subsystem in the network, by incorporating control signal $u_i$.}
    \label{Fig-conclusion}
\end{figure*}

What we have demonstrated across this review is how they offer different insight, but more importantly how they can be combined together. As illustrated in Figure \ref{Fig-conclusion}, we use opinion evolution formula to interlink the four methods. Topology-based centrality maximizes unbiased influence spread using graph information alone, represented by $A_{ij}$. In contrast, topic-sensitive methods rank users based on their ability to influence biased dynamics, using opinion states vector $x_i$ as the graph signal. By constructing opinion evolution models as $\dot x_i = f_i(x_i)+\sum_{j\neq i}^n A_{ij}g(x_i,x_j)$, we can apply graph sampling to identify orthogonal influencers, thereby minimizing redundant messaging. Further, by incorporating control signal $u_i$, we can determine which influencers maximally impact the controllable subsystem size.

In summary, our review highlights the diverse methodologies for identifying opinion leaders across disciplines. By integrating these different approaches, we can better understand the complex dynamics of opinion formation and influence in large networks.

\section{Conclusion}

Through this review and a case study, we performed a comparative analysis of multiple methodologies across disciplines. Some of the analysis we focus on is intra-disciplinary, showing connections and differences between graph centrality, control theory and sampling theory. Our initial contribution is cross-disciplinary in nature, reviewing qualitatively from the perspective of different disciplines. The greatest contribution is that we go forward and build connections to psychology and develop psychology-informed differential equation signals that can be combined with aforementioned graph signal analysis. This shows a trans-disciplinary contribution to knowledge.

The results show that a horizontal comparison among different ranking strategies is challenging, due to the disparate criteria utilised by the methods. There may be some overlap between the identified opinion leaders through various methods, yet their correlation and causality require further studies. It is our hope that this survey will help researches in gaining better understanding of the development of opinion leader detection methods and inspire them to address the remaining challenges in this field.

For future work, the main challenge would be integrating diverse contents, features and dynamics as input data to find the opinion leaders. Diverse contents refers to incorporation of cross-platform content analysis—for instance, text content on Twitter, graphical content on Instagram, and video content on TikTok. Various features can include user traits, content and network features. Dynamic nature of social network and human behavior presents another challenge. The continuous flow of data offers opportunities to identify emerging influencers who have the potential to influence public opinion. We also wish to consider how we can create synthetic test environments using emerging large language model agents \cite{R1}, which can create ethical environments to validate experiments. This can aid the development of larger diverse data sets with topic-specific influence labels is also important, as we have so far been limited to two widely used data sets.

Overall, a comprehensive approach is essential for identifying influential users using the high dimensional data resources. It is our hope that this survey will help researches in gaining better understanding of the development of opinion leader detection methods and inspire them to address the remaining challenges in this field.

\bmhead{Acknowledgments}

The work is supported by "Networked Social Influence and Acceptance in a New Age of Crises", funded by USAF OFSR under Grant No.: FA8655-20-1-7031, and is partly supported by the Engineering and Physical Sciences Research Council [grant number:  EP/V026763/1]

\section*{Statements and Declarations}

\begin{itemize}
\item Funding

The work is supported by "Networked Social Influence and Acceptance in a New Age of Crises", funded by USAF OFSR under Grant No.: FA8655-20-1-7031, and is partly supported by the Engineering and Physical Sciences Research Council [grant number:  EP/V026763/1]

\item Conflict of interest/Competing interests 

The authors have no relevant financial or non-financial interests to disclose.

\item Ethics approval and consent to participate

Not applicable.

\item Consent for publication

All authors of this paper have given their consent for its publication.

\item Data availability 

The data is published at https://github.com/AlminaJin/OpinionLeaderDetection.git.

\item Materials availability

Not applicable.

\item Code availability

The code is published at https://github.com/AlminaJin/OpinionLeaderDetection.git.

\item Author contribution

All authors contributed to the conception and design of the study. Bailu Jin was responsible for data collection. Bailu Jin, Mengbang Zou, and Zhuangkun Wei conducted the literature search, data analysis, and wrote the first draft of the manuscript. Weisi Guo supervised the project, and provided critical reviews and commentary on the work. All authors have read, revised, and approved the final manuscript.

\end{itemize}

\bibliography{Ref}


\begin{thebibliography}{65}
\ifx \bisbn   \undefined \def \bisbn  #1{ISBN #1}\fi
\ifx \binits  \undefined \def \binits#1{#1}\fi
\ifx \bauthor  \undefined \def \bauthor#1{#1}\fi
\ifx \batitle  \undefined \def \batitle#1{#1}\fi
\ifx \bjtitle  \undefined \def \bjtitle#1{#1}\fi
\ifx \bvolume  \undefined \def \bvolume#1{\textbf{#1}}\fi
\ifx \byear  \undefined \def \byear#1{#1}\fi
\ifx \bissue  \undefined \def \bissue#1{#1}\fi
\ifx \bfpage  \undefined \def \bfpage#1{#1}\fi
\ifx \blpage  \undefined \def \blpage #1{#1}\fi
\ifx \burl  \undefined \def \burl#1{\textsf{#1}}\fi
\ifx \doiurl  \undefined \def \doiurl#1{\url{https://doi.org/#1}}\fi
\ifx \betal  \undefined \def \betal{\textit{et al.}}\fi
\ifx \binstitute  \undefined \def \binstitute#1{#1}\fi
\ifx \binstitutionaled  \undefined \def \binstitutionaled#1{#1}\fi
\ifx \bctitle  \undefined \def \bctitle#1{#1}\fi
\ifx \beditor  \undefined \def \beditor#1{#1}\fi
\ifx \bpublisher  \undefined \def \bpublisher#1{#1}\fi
\ifx \bbtitle  \undefined \def \bbtitle#1{#1}\fi
\ifx \bedition  \undefined \def \bedition#1{#1}\fi
\ifx \bseriesno  \undefined \def \bseriesno#1{#1}\fi
\ifx \blocation  \undefined \def \blocation#1{#1}\fi
\ifx \bsertitle  \undefined \def \bsertitle#1{#1}\fi
\ifx \bsnm \undefined \def \bsnm#1{#1}\fi
\ifx \bsuffix \undefined \def \bsuffix#1{#1}\fi
\ifx \bparticle \undefined \def \bparticle#1{#1}\fi
\ifx \barticle \undefined \def \barticle#1{#1}\fi
\bibcommenthead
\ifx \bconfdate \undefined \def \bconfdate #1{#1}\fi
\ifx \botherref \undefined \def \botherref #1{#1}\fi
\ifx \url \undefined \def \url#1{\textsf{#1}}\fi
\ifx \bchapter \undefined \def \bchapter#1{#1}\fi
\ifx \bbook \undefined \def \bbook#1{#1}\fi
\ifx \bcomment \undefined \def \bcomment#1{#1}\fi
\ifx \oauthor \undefined \def \oauthor#1{#1}\fi
\ifx \citeauthoryear \undefined \def \citeauthoryear#1{#1}\fi
\ifx \endbibitem  \undefined \def \endbibitem {}\fi
\ifx \bconflocation  \undefined \def \bconflocation#1{#1}\fi
\ifx \arxivurl  \undefined \def \arxivurl#1{\textsf{#1}}\fi
\csname PreBibitemsHook\endcsname

\bibitem[\protect\citeauthoryear{Lazarsfeld et~al.}{1968}]{lazarsfeld1968people}
\begin{botherref}
\oauthor{\bsnm{Lazarsfeld}, \binits{P.F.}},
\oauthor{\bsnm{Berelson}, \binits{B.}},
\oauthor{\bsnm{Gaudet}, \binits{H.}}:
The people’s choice.
Columbia University Press
(1968)
\end{botherref}
\endbibitem

\bibitem[\protect\citeauthoryear{Bamakan et~al.}{2019}]{bamakan2019opinion}
\begin{barticle}
\bauthor{\bsnm{Bamakan}, \binits{S.M.H.}},
\bauthor{\bsnm{Nurgaliev}, \binits{I.}},
\bauthor{\bsnm{Qu}, \binits{Q.}}:
\batitle{Opinion leader detection: A methodological review}.
\bjtitle{Expert Systems with Applications}
\bvolume{115},
\bfpage{200}--\blpage{222}
(\byear{2019})
\end{barticle}
\endbibitem

\bibitem[\protect\citeauthoryear{Riquelme and Gonz{\'a}lez-Cantergiani}{2016}]{riquelme2016measuring}
\begin{barticle}
\bauthor{\bsnm{Riquelme}, \binits{F.}},
\bauthor{\bsnm{Gonz{\'a}lez-Cantergiani}, \binits{P.}}:
\batitle{Measuring user influence on twitter: A survey}.
\bjtitle{Information processing \& management}
\bvolume{52}(\bissue{5}),
\bfpage{949}--\blpage{975}
(\byear{2016})
\end{barticle}
\endbibitem

\bibitem[\protect\citeauthoryear{Panchendrarajan and Saxena}{2023}]{panchendrarajan2023topic}
\begin{barticle}
\bauthor{\bsnm{Panchendrarajan}, \binits{R.}},
\bauthor{\bsnm{Saxena}, \binits{A.}}:
\batitle{Topic-based influential user detection: a survey}.
\bjtitle{Applied Intelligence}
\bvolume{53}(\bissue{5}),
\bfpage{5998}--\blpage{6024}
(\byear{2023})
\end{barticle}
\endbibitem

\bibitem[\protect\citeauthoryear{Katz}{1957}]{katz1957two}
\begin{barticle}
\bauthor{\bsnm{Katz}, \binits{E.}}:
\batitle{The two-step flow of communication: An up-to-date report on an hypothesis}.
\bjtitle{Public opinion quarterly}
\bvolume{21}(\bissue{1}),
\bfpage{61}--\blpage{78}
(\byear{1957})
\end{barticle}
\endbibitem

\bibitem[\protect\citeauthoryear{Hellevik and Bj{\o}rklund}{1991}]{hellevik1991opinion}
\begin{barticle}
\bauthor{\bsnm{Hellevik}, \binits{O.}},
\bauthor{\bsnm{Bj{\o}rklund}, \binits{T.}}:
\batitle{Opinion leadership and political extremism}.
\bjtitle{International Journal of Public Opinion Research}
\bvolume{3}(\bissue{2}),
\bfpage{157}--\blpage{181}
(\byear{1991})
\end{barticle}
\endbibitem

\bibitem[\protect\citeauthoryear{Rogers and Cartano}{1962}]{rogers1962methods}
\begin{botherref}
\oauthor{\bsnm{Rogers}, \binits{E.M.}},
\oauthor{\bsnm{Cartano}, \binits{D.G.}}:
Methods of measuring opinion leadership.
Public opinion quarterly,
435--441
(1962)
\end{botherref}
\endbibitem

\bibitem[\protect\citeauthoryear{Corey}{1971}]{corey1971people}
\begin{barticle}
\bauthor{\bsnm{Corey}, \binits{L.G.}}:
\batitle{People who claim to be opinion leaders: identifying their characteristics by self-report}.
\bjtitle{Journal of Marketing}
\bvolume{35}(\bissue{4}),
\bfpage{48}--\blpage{53}
(\byear{1971})
\end{barticle}
\endbibitem

\bibitem[\protect\citeauthoryear{Tkachenko and Guo}{2020}]{10.1145/3297662.3365819}
\begin{bchapter}
\bauthor{\bsnm{Tkachenko}, \binits{N.}},
\bauthor{\bsnm{Guo}, \binits{W.}}:
\bctitle{Conflict detection in linguistically diverse on-line social networks: A russia-ukraine case study}.
In: \bbtitle{ACM International Conference on Management of Digital EcoSystems},
pp. \bfpage{23}--\blpage{28}
(\byear{2020})
\end{bchapter}
\endbibitem

\bibitem[\protect\citeauthoryear{French~Jr}{1956}]{french1956formal}
\begin{barticle}
\bauthor{\bsnm{French~Jr}, \binits{J.R.}}:
\batitle{A formal theory of social power.}
\bjtitle{Psychological review}
\bvolume{63}(\bissue{3}),
\bfpage{181}
(\byear{1956})
\end{barticle}
\endbibitem

\bibitem[\protect\citeauthoryear{Myers}{1982}]{myers1982polarizing}
\begin{barticle}
\bauthor{\bsnm{Myers}, \binits{D.G.}}:
\batitle{Polarizing effects of social interaction}.
\bjtitle{Group decision making}
\bvolume{125},
\bfpage{137}--\blpage{138}
(\byear{1982})
\end{barticle}
\endbibitem

\bibitem[\protect\citeauthoryear{DeGroot}{1974}]{degroot1974reaching}
\begin{barticle}
\bauthor{\bsnm{DeGroot}, \binits{M.H.}}:
\batitle{Reaching a consensus}.
\bjtitle{Journal of the American Statistical Association}
\bvolume{69}(\bissue{345}),
\bfpage{118}--\blpage{121}
(\byear{1974})
\end{barticle}
\endbibitem

\bibitem[\protect\citeauthoryear{Dong et~al.}{2017}]{R5}
\begin{botherref}
\oauthor{\bsnm{Dong}, \binits{Y.}},
\oauthor{\bsnm{Ding}, \binits{Z.}},
\oauthor{\bsnm{Martínez}, \binits{L.}},
\oauthor{\bsnm{Herrera}, \binits{F.}}:
Managing consensus based on leadership in opinion dynamics.
Information Sciences
\textbf{397}
(2017)
\end{botherref}
\endbibitem

\bibitem[\protect\citeauthoryear{Hegselmann et~al.}{2002}]{hegselmann2002opinion}
\begin{botherref}
\oauthor{\bsnm{Hegselmann}, \binits{R.}},
\oauthor{\bsnm{Krause}, \binits{U.}}, et al.:
Opinion dynamics and bounded confidence models, analysis, and simulation.
Journal of artificial societies and social simulation
\textbf{5}(3)
(2002)
\end{botherref}
\endbibitem

\bibitem[\protect\citeauthoryear{Bavelas}{1948}]{bavelas1948mathematical}
\begin{barticle}
\bauthor{\bsnm{Bavelas}, \binits{A.}}:
\batitle{A mathematical model for group structures}.
\bjtitle{Human organization}
\bvolume{7}(\bissue{3}),
\bfpage{16}--\blpage{30}
(\byear{1948})
\end{barticle}
\endbibitem

\bibitem[\protect\citeauthoryear{Freeman et~al.}{1979}]{freeman1979centrality}
\begin{barticle}
\bauthor{\bsnm{Freeman}, \binits{L.C.}}, \betal:
\batitle{Centrality in social networks: Conceptual clarification}.
\bjtitle{Social network: critical concepts in sociology. Londres: Routledge}
\bvolume{1},
\bfpage{215}--\blpage{239}
(\byear{1979})
\end{barticle}
\endbibitem

\bibitem[\protect\citeauthoryear{Yang et~al.}{2018}]{yang2018identifying}
\begin{barticle}
\bauthor{\bsnm{Yang}, \binits{L.}},
\bauthor{\bsnm{Qiao}, \binits{Y.}},
\bauthor{\bsnm{Liu}, \binits{Z.}},
\bauthor{\bsnm{Ma}, \binits{J.}},
\bauthor{\bsnm{Li}, \binits{X.}}:
\batitle{Identifying opinion leader nodes in online social networks with a new closeness evaluation algorithm}.
\bjtitle{Soft Computing}
\bvolume{22},
\bfpage{453}--\blpage{464}
(\byear{2018})
\end{barticle}
\endbibitem

\bibitem[\protect\citeauthoryear{Katz}{1953}]{katz1953new}
\begin{barticle}
\bauthor{\bsnm{Katz}, \binits{L.}}:
\batitle{A new status index derived from sociometric analysis}.
\bjtitle{Psychometrika}
\bvolume{18}(\bissue{1}),
\bfpage{39}--\blpage{43}
(\byear{1953})
\end{barticle}
\endbibitem

\bibitem[\protect\citeauthoryear{Page et~al.}{1999}]{page1999pagerank}
\begin{botherref}
\oauthor{\bsnm{Page}, \binits{L.}},
\oauthor{\bsnm{Brin}, \binits{S.}},
\oauthor{\bsnm{Motwani}, \binits{R.}},
\oauthor{\bsnm{Winograd}, \binits{T.}}:
The pagerank citation ranking: Bringing order to the web.
Technical report,
Stanford InfoLab
(1999)
\end{botherref}
\endbibitem

\bibitem[\protect\citeauthoryear{Kleinberg}{1999}]{kleinberg1999authoritative}
\begin{barticle}
\bauthor{\bsnm{Kleinberg}, \binits{J.M.}}:
\batitle{Authoritative sources in a hyperlinked environment}.
\bjtitle{Journal of the ACM (JACM)}
\bvolume{46}(\bissue{5}),
\bfpage{604}--\blpage{632}
(\byear{1999})
\end{barticle}
\endbibitem

\bibitem[\protect\citeauthoryear{Yeung et~al.}{2011}]{yeung2011spear}
\begin{barticle}
\bauthor{\bsnm{Yeung}, \binits{C.-m.A.}},
\bauthor{\bsnm{Noll}, \binits{M.G.}},
\bauthor{\bsnm{Gibbins}, \binits{N.}},
\bauthor{\bsnm{Meinel}, \binits{C.}},
\bauthor{\bsnm{Shadbolt}, \binits{N.}}:
\batitle{Spear: spamming-resistant expertise analysis and ranking in collaborative tagging systems}.
\bjtitle{Computational Intelligence}
\bvolume{27}(\bissue{3}),
\bfpage{458}--\blpage{488}
(\byear{2011})
\end{barticle}
\endbibitem

\bibitem[\protect\citeauthoryear{Shinde and Girase}{2016}]{shinde2016identification}
\begin{bchapter}
\bauthor{\bsnm{Shinde}, \binits{M.}},
\bauthor{\bsnm{Girase}, \binits{S.}}:
\bctitle{Identification of topic-specific opinion leader using spear algorithm in online knowledge communities}.
In: \bbtitle{2016 International Conference on Computing, Analytics and Security Trends (CAST)},
pp. \bfpage{144}--\blpage{149}
(\byear{2016}).
\bcomment{IEEE}
\end{bchapter}
\endbibitem

\bibitem[\protect\citeauthoryear{Tunkelang}{2009}]{tunkelang2009a}
\begin{botherref}
\oauthor{\bsnm{Tunkelang}, \binits{D.}}:
A twitter analog to pagerank
(2009)
\end{botherref}
\endbibitem

\bibitem[\protect\citeauthoryear{Kempe et~al.}{2003}]{kempe2003maximizing}
\begin{bchapter}
\bauthor{\bsnm{Kempe}, \binits{D.}},
\bauthor{\bsnm{Kleinberg}, \binits{J.}},
\bauthor{\bsnm{Tardos}, \binits{{\'E}.}}:
\bctitle{Maximizing the spread of influence through a social network}.
In: \bbtitle{Proceedings of the Ninth ACM SIGKDD International Conference on Knowledge Discovery and Data Mining},
pp. \bfpage{137}--\blpage{146}
(\byear{2003})
\end{bchapter}
\endbibitem

\bibitem[\protect\citeauthoryear{Zhao et~al.}{2015}]{zhao2015seismic}
\begin{bchapter}
\bauthor{\bsnm{Zhao}, \binits{Q.}},
\bauthor{\bsnm{Erdogdu}, \binits{M.A.}},
\bauthor{\bsnm{He}, \binits{H.Y.}},
\bauthor{\bsnm{Rajaraman}, \binits{A.}},
\bauthor{\bsnm{Leskovec}, \binits{J.}}:
\bctitle{Seismic: A self-exciting point process model for predicting tweet popularity}.
In: \bbtitle{Proceedings of the 21th ACM SIGKDD International Conference on Knowledge Discovery and Data Mining},
pp. \bfpage{1513}--\blpage{1522}
(\byear{2015})
\end{bchapter}
\endbibitem

\bibitem[\protect\citeauthoryear{Wang et~al.}{2024}]{R2}
\begin{botherref}
\oauthor{\bsnm{Wang}, \binits{Y.}},
\oauthor{\bsnm{Li}, \binits{H.}},
\oauthor{\bsnm{Zhang}, \binits{L.}},
\oauthor{\bsnm{Zhao}, \binits{L.}},
\oauthor{\bsnm{Li}, \binits{W.}}:
Identifying influential nodes in social networks: Centripetal centrality and seed exclusion approach.
Chaos
\textbf{163}
(2024)
\end{botherref}
\endbibitem

\bibitem[\protect\citeauthoryear{Rashid and Bhat}{2023}]{R3}
\begin{botherref}
\oauthor{\bsnm{Rashid}, \binits{Y.}},
\oauthor{\bsnm{Bhat}, \binits{J.}}:
Topological to deep learning era for identifying influencers in online social networks: a systematic review.
Multimedia Tools \& Applications
\textbf{83}
(2023)
\end{botherref}
\endbibitem

\bibitem[\protect\citeauthoryear{Zhou et~al.}{2024}]{R4}
\begin{botherref}
\oauthor{\bsnm{Zhou}, \binits{F.}},
\oauthor{\bsnm{Lv}, \binits{L.}},
\oauthor{\bsnm{Liu}, \binits{J.}},
\oauthor{\bsnm{Mariani}, \binits{M.S.}}:
Beyond network centrality: individual-level behavioral traits for predicting information superspreaders in social media.
National Science Review
\textbf{11}
(2024)
\end{botherref}
\endbibitem

\bibitem[\protect\citeauthoryear{Cha et~al.}{2010}]{cha2010measuring}
\begin{bchapter}
\bauthor{\bsnm{Cha}, \binits{M.}},
\bauthor{\bsnm{Haddadi}, \binits{H.}},
\bauthor{\bsnm{Benevenuto}, \binits{F.}}:
\bctitle{Measuring user influence in twitter: The million follower fallacy}.
In: \bbtitle{Proceedings Of The Fourth International Aaai Conference On Weblogs And Social Media}
(\byear{2010})
\end{bchapter}
\endbibitem

\bibitem[\protect\citeauthoryear{Song et~al.}{2007}]{song2007identifying}
\begin{bchapter}
\bauthor{\bsnm{Song}, \binits{X.}},
\bauthor{\bsnm{Chi}, \binits{Y.}},
\bauthor{\bsnm{Hino}, \binits{K.}},
\bauthor{\bsnm{Tseng}, \binits{B.}}:
\bctitle{Identifying opinion leaders in the blogosphere}.
In: \bbtitle{Proceedings of the Sixteenth ACM Conference on Conference on Information and Knowledge Management},
pp. \bfpage{971}--\blpage{974}
(\byear{2007})
\end{bchapter}
\endbibitem

\bibitem[\protect\citeauthoryear{McPherson et~al.}{2001}]{mcpherson2001birds}
\begin{barticle}
\bauthor{\bsnm{McPherson}, \binits{M.}},
\bauthor{\bsnm{Smith-Lovin}, \binits{L.}},
\bauthor{\bsnm{Cook}, \binits{J.M.}}:
\batitle{Birds of a feather: Homophily in social networks}.
\bjtitle{Annual review of sociology}
\bvolume{27}(\bissue{1}),
\bfpage{415}--\blpage{444}
(\byear{2001})
\end{barticle}
\endbibitem

\bibitem[\protect\citeauthoryear{Weng et~al.}{2010}]{weng2010twitterrank}
\begin{bchapter}
\bauthor{\bsnm{Weng}, \binits{J.}},
\bauthor{\bsnm{Lim}, \binits{E.-P.}},
\bauthor{\bsnm{Jiang}, \binits{J.}},
\bauthor{\bsnm{He}, \binits{Q.}}:
\bctitle{Twitterrank: finding topic-sensitive influential twitterers}.
In: \bbtitle{Proceedings of the Third ACM International Conference on Web Search and Data Mining},
pp. \bfpage{261}--\blpage{270}
(\byear{2010})
\end{bchapter}
\endbibitem

\bibitem[\protect\citeauthoryear{Dhali et~al.}{2020}]{dhali2020attribute}
\begin{bchapter}
\bauthor{\bsnm{Dhali}, \binits{A.}},
\bauthor{\bsnm{Gomasta}, \binits{S.S.}},
\bauthor{\bsnm{Anwar}, \binits{M.M.}},
\bauthor{\bsnm{Sarker}, \binits{I.H.}}:
\bctitle{Attribute-driven topical influential users detection in online social networks}.
In: \bbtitle{2020 IEEE Asia-Pacific Conference on Computer Science and Data Engineering (CSDE)},
pp. \bfpage{1}--\blpage{5}
(\byear{2020}).
\bcomment{IEEE}
\end{bchapter}
\endbibitem

\bibitem[\protect\citeauthoryear{Eliacik and Erdogan}{2018}]{eliacik2018influential}
\begin{barticle}
\bauthor{\bsnm{Eliacik}, \binits{A.B.}},
\bauthor{\bsnm{Erdogan}, \binits{N.}}:
\batitle{Influential user weighted sentiment analysis on topic based microblogging community}.
\bjtitle{Expert Systems with Applications}
\bvolume{92},
\bfpage{403}--\blpage{418}
(\byear{2018})
\end{barticle}
\endbibitem

\bibitem[\protect\citeauthoryear{Pal and Counts}{2011}]{pal2011identifying}
\begin{bchapter}
\bauthor{\bsnm{Pal}, \binits{A.}},
\bauthor{\bsnm{Counts}, \binits{S.}}:
\bctitle{Identifying topical authorities in microblogs}.
In: \bbtitle{Proceedings of the Fourth ACM International Conference on Web Search and Data Mining},
pp. \bfpage{45}--\blpage{54}
(\byear{2011})
\end{bchapter}
\endbibitem

\bibitem[\protect\citeauthoryear{Zhou et~al.}{2009}]{zhou2009finding}
\begin{bchapter}
\bauthor{\bsnm{Zhou}, \binits{H.}},
\bauthor{\bsnm{Zeng}, \binits{D.}},
\bauthor{\bsnm{Zhang}, \binits{C.}}:
\bctitle{Finding leaders from opinion networks}.
In: \bbtitle{2009 IEEE International Conference on Intelligence and Security Informatics},
pp. \bfpage{266}--\blpage{268}
(\byear{2009}).
\bcomment{IEEE}
\end{bchapter}
\endbibitem

\bibitem[\protect\citeauthoryear{Huang et~al.}{2014}]{huang2014finding}
\begin{botherref}
\oauthor{\bsnm{Huang}, \binits{B.}},
\oauthor{\bsnm{Yu}, \binits{G.}},
\oauthor{\bsnm{Karimi}, \binits{H.R.}}:
The finding and dynamic detection of opinion leaders in social network.
Mathematical problems in engineering
\textbf{2014}
(2014)
\end{botherref}
\endbibitem

\bibitem[\protect\citeauthoryear{Dong et~al.}{2019}]{dong2019mining}
\begin{barticle}
\bauthor{\bsnm{Dong}, \binits{G.}},
\bauthor{\bsnm{Li}, \binits{B.}},
\bauthor{\bsnm{Wei}, \binits{X.}},
\bauthor{\bsnm{Qin}, \binits{T.}}:
\batitle{Mining key users of microblog topics based on trust model}.
\bjtitle{International Journal of Performability Engineering}
\bvolume{15}(\bissue{11}),
\bfpage{3024}
(\byear{2019})
\end{barticle}
\endbibitem

\bibitem[\protect\citeauthoryear{Chen et~al.}{2014}]{chen2014identifying}
\begin{bchapter}
\bauthor{\bsnm{Chen}, \binits{Y.}},
\bauthor{\bsnm{Wang}, \binits{X.}},
\bauthor{\bsnm{Tang}, \binits{B.}},
\bauthor{\bsnm{Xu}, \binits{R.}},
\bauthor{\bsnm{Yuan}, \binits{B.}},
\bauthor{\bsnm{Xiang}, \binits{X.}},
\bauthor{\bsnm{Bu}, \binits{J.}}:
\bctitle{Identifying opinion leaders from online comments}.
In: \bbtitle{Social Media Processing: Third National Conference, SMP 2014, Beijing, China, November 1-2, 2014. Proceedings},
pp. \bfpage{231}--\blpage{239}
(\byear{2014}).
\bcomment{Springer}
\end{bchapter}
\endbibitem

\bibitem[\protect\citeauthoryear{Jin and Guo}{2022}]{10068693}
\begin{bchapter}
\bauthor{\bsnm{Jin}, \binits{B.}},
\bauthor{\bsnm{Guo}, \binits{W.}}:
\bctitle{Data driven modeling social media influence using differential equations}.
In: \bbtitle{2022 IEEE/ACM International Conference on Advances in Social Networks Analysis and Mining (ASONAM)},
pp. \bfpage{504}--\blpage{507}
(\byear{2022})
\end{bchapter}
\endbibitem

\bibitem[\protect\citeauthoryear{Cercel and Trausan-Matu}{2014}]{cercel2014opinion}
\begin{bchapter}
\bauthor{\bsnm{Cercel}, \binits{D.-C.}},
\bauthor{\bsnm{Trausan-Matu}, \binits{S.}}:
\bctitle{Opinion propagation in online social networks: A survey}.
In: \bbtitle{Proceedings of the 4th International Conference on Web Intelligence, Mining and Semantics (WIMS14)},
pp. \bfpage{1}--\blpage{10}
(\byear{2014})
\end{bchapter}
\endbibitem

\bibitem[\protect\citeauthoryear{Kalman}{1963}]{kalman1963mathematical}
\begin{barticle}
\bauthor{\bsnm{Kalman}, \binits{R.E.}}:
\batitle{Mathematical description of linear dynamical systems}.
\bjtitle{Journal of the Society for Industrial and Applied Mathematics, Series A: Control}
\bvolume{1}(\bissue{2}),
\bfpage{152}--\blpage{192}
(\byear{1963})
\end{barticle}
\endbibitem

\bibitem[\protect\citeauthoryear{Lin}{1974}]{lin1974structural}
\begin{barticle}
\bauthor{\bsnm{Lin}, \binits{C.-T.}}:
\batitle{Structural controllability}.
\bjtitle{IEEE Transactions on Automatic Control}
\bvolume{19}(\bissue{3}),
\bfpage{201}--\blpage{208}
(\byear{1974})
\end{barticle}
\endbibitem

\bibitem[\protect\citeauthoryear{Liu and Barab{\'a}si}{2016}]{liu2016control}
\begin{barticle}
\bauthor{\bsnm{Liu}, \binits{Y.-Y.}},
\bauthor{\bsnm{Barab{\'a}si}, \binits{A.-L.}}:
\batitle{Control principles of complex systems}.
\bjtitle{Reviews of Modern Physics}
\bvolume{88}(\bissue{3}),
\bfpage{035006}
(\byear{2016})
\end{barticle}
\endbibitem

\bibitem[\protect\citeauthoryear{Murota}{2010}]{murota2010matrices}
\begin{bbook}
\bauthor{\bsnm{Murota}, \binits{K.}}:
\bbtitle{Matrices and Matroids for Systems Analysis}.
\bpublisher{Springer}, \blocation{???}
(\byear{2010})
\end{bbook}
\endbibitem

\bibitem[\protect\citeauthoryear{Liu et~al.}{2011}]{liu2011controllability}
\begin{barticle}
\bauthor{\bsnm{Liu}, \binits{Y.-Y.}},
\bauthor{\bsnm{Slotine}, \binits{J.-J.}},
\bauthor{\bsnm{Barab{\'a}si}, \binits{A.-L.}}:
\batitle{Controllability of complex networks}.
\bjtitle{nature}
\bvolume{473}(\bissue{7346}),
\bfpage{167}--\blpage{173}
(\byear{2011})
\end{barticle}
\endbibitem

\bibitem[\protect\citeauthoryear{Liu et~al.}{2012}]{liu2012control}
\begin{botherref}
\oauthor{\bsnm{Liu}, \binits{Y.-Y.}},
\oauthor{\bsnm{Slotine}, \binits{J.-J.}},
\oauthor{\bsnm{Barab{\'a}si}, \binits{A.-L.}}:
Control centrality and hierarchical structure in complex networks
(2012)
\end{botherref}
\endbibitem

\bibitem[\protect\citeauthoryear{Hosoe}{1980}]{hosoe1980determination}
\begin{barticle}
\bauthor{\bsnm{Hosoe}, \binits{S.}}:
\batitle{Determination of generic dimensions of controllable subspaces and its application}.
\bjtitle{IEEE Transactions on Automatic Control}
\bvolume{25}(\bissue{6}),
\bfpage{1192}--\blpage{1196}
(\byear{1980})
\end{barticle}
\endbibitem

\bibitem[\protect\citeauthoryear{Zou et~al.}{2024}]{10552434}
\begin{barticle}
\bauthor{\bsnm{Zou}, \binits{M.}},
\bauthor{\bsnm{Guo}, \binits{W.}},
\bauthor{\bsnm{Chu}, \binits{K.-F.}}:
\batitle{Rewiring complex networks to achieve cluster synchronization using graph convolution networks with reinforcement learning}.
\bjtitle{IEEE Transactions on Network Science and Engineering}
\bvolume{11}(\bissue{5}),
\bfpage{4293}--\blpage{4304}
(\byear{2024})
\end{barticle}
\endbibitem

\bibitem[\protect\citeauthoryear{Jing et~al.}{2018}]{8233154}
\begin{barticle}
\bauthor{\bsnm{Jing}, \binits{P.}},
\bauthor{\bsnm{Su}, \binits{Y.}},
\bauthor{\bsnm{Nie}, \binits{L.}},
\bauthor{\bsnm{Bai}, \binits{X.}},
\bauthor{\bsnm{Liu}, \binits{J.}},
\bauthor{\bsnm{Wang}, \binits{M.}}:
\batitle{Low-rank multi-view embedding learning for micro-video popularity prediction}.
\bjtitle{IEEE Transactions on Knowledge and Data Engineering}
\bvolume{30}(\bissue{8}),
\bfpage{1519}--\blpage{1532}
(\byear{2018})
\end{barticle}
\endbibitem

\bibitem[\protect\citeauthoryear{Zuo et~al.}{2018}]{8292837}
\begin{barticle}
\bauthor{\bsnm{Zuo}, \binits{X.}},
\bauthor{\bsnm{Liu}, \binits{X.}},
\bauthor{\bsnm{Yang}, \binits{B.}}:
\batitle{Coupled low rank approximation for collaborative filtering in social networks}.
\bjtitle{IEEE Access}
\bvolume{6},
\bfpage{13326}--\blpage{13335}
(\byear{2018})
\end{barticle}
\endbibitem

\bibitem[\protect\citeauthoryear{Donavalli et~al.}{2012}]{donavalli2012low}
\begin{bchapter}
\bauthor{\bsnm{Donavalli}, \binits{A.}},
\bauthor{\bsnm{Rege}, \binits{M.}},
\bauthor{\bsnm{Liu}, \binits{X.}},
\bauthor{\bsnm{Jafari-Khouzani}, \binits{K.}}:
\bctitle{Low-rank matrix factorization and co-clustering algorithms for analyzing large data sets}.
In: \bbtitle{Data Engineering and Management: Second International Conference, ICDEM 2010, Tiruchirappalli, India, July 29-31, 2010. Revised Selected Papers},
pp. \bfpage{272}--\blpage{279}
(\byear{2012}).
\bcomment{Springer}
\end{bchapter}
\endbibitem

\bibitem[\protect\citeauthoryear{Candes and Recht}{2012}]{candes2012exact}
\begin{barticle}
\bauthor{\bsnm{Candes}, \binits{E.}},
\bauthor{\bsnm{Recht}, \binits{B.}}:
\batitle{Exact matrix completion via convex optimization}.
\bjtitle{Communications of the ACM}
\bvolume{55}(\bissue{6}),
\bfpage{111}--\blpage{119}
(\byear{2012})
\end{barticle}
\endbibitem

\bibitem[\protect\citeauthoryear{Mahoney and Drineas}{2009}]{mahoney2009cur}
\begin{barticle}
\bauthor{\bsnm{Mahoney}, \binits{M.W.}},
\bauthor{\bsnm{Drineas}, \binits{P.}}:
\batitle{Cur matrix decompositions for improved data analysis}.
\bjtitle{Proceedings of the National Academy of Sciences}
\bvolume{106}(\bissue{3}),
\bfpage{697}--\blpage{702}
(\byear{2009})
\end{barticle}
\endbibitem

\bibitem[\protect\citeauthoryear{Qiu et~al.}{2017}]{7979523}
\begin{barticle}
\bauthor{\bsnm{Qiu}, \binits{K.}},
\bauthor{\bsnm{Mao}, \binits{X.}},
\bauthor{\bsnm{Shen}, \binits{X.}},
\bauthor{\bsnm{Wang}, \binits{X.}},
\bauthor{\bsnm{Li}, \binits{T.}},
\bauthor{\bsnm{Gu}, \binits{Y.}}:
\batitle{Time-varying graph signal reconstruction}.
\bjtitle{IEEE Journal of Selected Topics in Signal Processing}
\bvolume{11}(\bissue{6}),
\bfpage{870}--\blpage{883}
(\byear{2017})
\end{barticle}
\endbibitem

\bibitem[\protect\citeauthoryear{Wei et~al.}{2020a}]{9234614}
\begin{barticle}
\bauthor{\bsnm{Wei}, \binits{Z.}},
\bauthor{\bsnm{Li}, \binits{B.}},
\bauthor{\bsnm{Sun}, \binits{C.}},
\bauthor{\bsnm{Guo}, \binits{W.}}:
\batitle{Sampling and inference of networked dynamics using log-koopman nonlinear graph fourier transform}.
\bjtitle{IEEE Transactions on Signal Processing}
\bvolume{68},
\bfpage{6187}--\blpage{6197}
(\byear{2020})
\end{barticle}
\endbibitem

\bibitem[\protect\citeauthoryear{Wei et~al.}{2020b}]{8839864}
\begin{barticle}
\bauthor{\bsnm{Wei}, \binits{Z.}},
\bauthor{\bsnm{Pagani}, \binits{A.}},
\bauthor{\bsnm{Fu}, \binits{G.}},
\bauthor{\bsnm{Guymer}, \binits{I.}},
\bauthor{\bsnm{Chen}, \binits{W.}},
\bauthor{\bsnm{McCann}, \binits{J.}},
\bauthor{\bsnm{Guo}, \binits{W.}}:
\batitle{Optimal sampling of water distribution network dynamics using graph fourier transform}.
\bjtitle{IEEE Transactions on Network Science and Engineering}
\bvolume{7}(\bissue{3}),
\bfpage{1570}--\blpage{1582}
(\byear{2020})
\end{barticle}
\endbibitem

\bibitem[\protect\citeauthoryear{Wei et~al.}{2019a}]{8865055}
\begin{barticle}
\bauthor{\bsnm{Wei}, \binits{Z.}},
\bauthor{\bsnm{Li}, \binits{B.}},
\bauthor{\bsnm{Guo}, \binits{W.}}:
\batitle{Optimal sampling for dynamic complex networks with graph-bandlimited initialization}.
\bjtitle{IEEE Access}
\bvolume{7},
\bfpage{150294}--\blpage{150305}
(\byear{2019})
\end{barticle}
\endbibitem

\bibitem[\protect\citeauthoryear{Wei et~al.}{2019b}]{9071735}
\begin{bchapter}
\bauthor{\bsnm{Wei}, \binits{Z.}},
\bauthor{\bsnm{Pagani}, \binits{A.}},
\bauthor{\bsnm{Guo}, \binits{W.}}:
\bctitle{Monitoring networked infrastructure with minimum data via sequential graph fourier transforms}.
In: \bbtitle{2019 IEEE International Smart Cities Conference (ISC2)},
pp. \bfpage{703}--\blpage{708}
(\byear{2019})
\end{bchapter}
\endbibitem

\bibitem[\protect\citeauthoryear{Dey}{2022}]{dey2022dynamic}
\begin{botherref}
\oauthor{\bsnm{Dey}, \binits{S.}}:
Dynamic mode decomposition and koopman theory.
arXiv preprint arXiv:2211.07561
(2022)
\end{botherref}
\endbibitem

\bibitem[\protect\citeauthoryear{Brunton}{2019}]{brunton2019notes}
\begin{botherref}
\oauthor{\bsnm{Brunton}, \binits{S.L.}}:
Notes on koopman operator theory.
Universit{\"a}t von Washington, Department of Mechanical Engineering, Zugriff
\textbf{30}
(2019)
\end{botherref}
\endbibitem

\bibitem[\protect\citeauthoryear{Valente and Pumpuang}{2007}]{valente2007identifying}
\begin{barticle}
\bauthor{\bsnm{Valente}, \binits{T.W.}},
\bauthor{\bsnm{Pumpuang}, \binits{P.}}:
\batitle{Identifying opinion leaders to promote behavior change}.
\bjtitle{Health education \& behavior}
\bvolume{34}(\bissue{6}),
\bfpage{881}--\blpage{896}
(\byear{2007})
\end{barticle}
\endbibitem

\bibitem[\protect\citeauthoryear{Bruns and Stieglitz}{2014}]{bruns2014metrics}
\begin{botherref}
\oauthor{\bsnm{Bruns}, \binits{A.}},
\oauthor{\bsnm{Stieglitz}, \binits{S.}}:
Metrics for understanding communication on twitter.
Twitter and society [Digital Formations, Volume 89],
69--82
(2014)
\end{botherref}
\endbibitem

\bibitem[\protect\citeauthoryear{Rogers et~al.}{2014}]{rogers2014diffusion}
\begin{botherref}
\oauthor{\bsnm{Rogers}, \binits{E.M.}},
\oauthor{\bsnm{Singhal}, \binits{A.}},
\oauthor{\bsnm{Quinlan}, \binits{M.M.}}:
Diffusion of innovations.
Routledge,
432--448
(2014)
\end{botherref}
\endbibitem

\bibitem[\protect\citeauthoryear{Siahkali et~al.}{2024}]{R1}
\begin{botherref}
\oauthor{\bsnm{Siahkali}, \binits{F.}},
\oauthor{\bsnm{Samadi}, \binits{S.}},
\oauthor{\bsnm{Kebriaei}, \binits{H.}}:
Towards opinion shaping: A deep reinforcement learning approach in bot-user interactions.
arXiv:2409.11426
(2024)
\end{botherref}
\endbibitem

\end{thebibliography}

\end{document}